\documentclass[preprint]{aastex631}
\usepackage{graphicx}

\submitjournal{\apj}

\accepted{October 28, 2022}

\shorttitle{The Effect of Duration of the Shock-Compressed Layer on Star Formation}
\shortauthors{Abe et al.}

\begin{document}

\title{The Effect of Shock Wave Duration on Star Formation and the Initial Condition of Massive Cluster Formation}

\email{d.abe@nagoya-u.jp}

\author[0000-0001-6891-2995]{Daisei Abe}
\affiliation{Department of Physics, Graduate School of Science, Nagoya University, Furo-cho, Chikusa-ku, Nagoya 464-8602, Japan}

\author[0000-0002-7935-8771]{Tsuyoshi Inoue}
\affiliation{Department of Physics, Faculty of Science
and Engineering, Konan University, Okamoto 8-9-1, Higashinada-ku, Kobe 658-8501, Japan}
\affiliation{Department of Physics, Graduate School of Science, Nagoya University, Furo-cho, Chikusa-ku, Nagoya 464-8602, Japan}

\author[0000-0003-2735-3239]{Rei Enokiya}
\affiliation{Department of Physics, Faculty of Science and Technology, Keio University, 3-14-1 Hiyoshi, Kohoku-ku, Yokohama, Kanagawa 223-8522, Japan}

\author[0000-0002-8966-9856]{Yasuo Fukui}
\affiliation{Department of Physics, Graduate School of Science, Nagoya University, Furo-cho, Chikusa-ku, Nagoya 464-8602, Japan}

\graphicspath{{./}{figures/}}

\begin{abstract}
Stars are born in dense molecular filaments irrespective of their mass.
{Compression of the ISM by shocks} cause filament formation in molecular clouds.
Observations show that a massive star cluster formation occurs where the peak of gas column density in a cloud exceeds 10$^{23}$ cm$^{-2}$.
In this study, we investigate the effect of the shock-compressed layer duration on filament/star formation and how the initial conditions of massive star formation are realized by performing three-dimensional (3D) isothermal magnetohydrodynamics (MHD) simulations with {gas inflow duration from the boundaries (i.e., shock wave duration)} as a controlling parameter.
Filaments formed behind the shock expand after the duration time for short shock duration models, whereas long duration models lead to star formation by forming massive supercritical filaments.
Moreover, when the shock duration is longer than two postshock free-fall times, the peak column density of the compressed layer exceeds 10$^{23}$ cm$^{-2}$, and {the gravitational collapse of the layer causes that} the number of OB stars expected to be formed in the shock-compressed layer reaches the order of ten (i.e., massive cluster formation).
\end{abstract}


\section{Introduction} \label{sec:intro}
Revealing the triggering mechanism of massive star/cluster formation is a critical target because the massive stars strongly affect the dynamics and evolution of galaxies.
Observations show that stars are born in dense molecular filaments irrespective of their mass~\citep[][]{andre2010A&A...518L.102A,fukui2015ApJ...807L...4F,fukui2019ApJ...886...14F,shimajiri2019A&A...632A..83S}.
The formation of filaments has been extensively studied~\citep[e.g., ][]{Tomisaka1983,Nagai1998,Padoan1999,Hennebelle2013,Pudritz2013RSPTA.37120248P,inoue2013ApJ...774L..31I,chenOstriker2014ApJ...785...69C,Balfour2015,Balfour2017,Federrath2016,abe2021ApJ...916...83A}.
Recently, \cite{abe2021ApJ...916...83A} found that the major filament formation mechanism changes with the shock velocity triggering the filament formation.
One of the major mechanisms is ``Type O"~\citep{inoue2013ApJ...774L..31I,Vaidya2013,abe2021ApJ...916...83A}, the details of which is as follows:
When the shock wave sweeps a dense clump in the cloud, the shock surface deforms due to large inertia of the clump.
Because the deformed shock wave bends the streamlines across the shock, the gas flows toward the convex point of the deformed shock wave, where only the component of flows parallel to the post-shock magnetic field lines converge and {lead} to the formation of a dense filament.
A high Mach number shock is required to induce massive star formation~\citep[e.g.,][]{Fukui2021,Liow2020MNRAS.499.1099L,Dobbs2020MNRAS.496L...1D}, and a massive filament is formed in the shock-compressed layer by Type O mechanism~\citep{abe2021ApJ...916...83A}.

Several theoretical studies used numerical simulations of shock compression of molecular clouds via cloud collision.
Those results were compared with observations~\citep[][]{Habe1992PASJ...44..203H,Takahira2014ApJ...792...63T,Bisbas2017ApJ...850...23B,Wu2017ApJ...835..137W,Wu2017ApJ...841...88W,Liow2020MNRAS.499.1099L,Dobbs2020MNRAS.496L...1D,Sakre2021PASJ...73S.385S}.
Previous studies have demonstrated that strong MHD shock compression allows for the formation of massive cores~\citep[][]{inoue2013ApJ...774L..31I,Inoue2018PASJ...70S..53I,Sakre2021PASJ...73S.385S}.
However, these studies implicitly assumed a long duration of a shock wave (corresponding {to} collisions between large clouds), whereas the realistic duration of a shock wave depends on the situation, e.g., the shock created by cloud collision cannot be kept over a crossing time.
On the other hand, simulations of unmagnetized cloud collisions by \citet[][]{Takahira2014ApJ...792...63T} demonstrated that gravitationally bound cores do not form in the case of short duration.
Therefore, a systematic study into the effect of shock duration on the resulting filament/core formation is required.

\citet[][]{enokiya2021PASJ...73S..75E} demonstrated that the peak column density of a star-forming region correlates with the number of OB stars in the system, and massive star clusters with more than 10 OB-type stars are associated with massive clouds whose peak column density exceeds 10$^{23}$ cm$^{-2}$.
{Thus}, the physical origin of the threshold peak column density of 10$^{23}$ cm$^{-2}$ must be clarified.
{It should be noted that the mean column density $\bar{N}_{\mathrm{H}_2}$ is related to the shock duration $t_{\mathrm{dur}}$, because $\bar{N}_{\mathrm{H}_2}=nL$ and $t_{\mathrm{dur}}=L/v_{\mathrm{sh}}$ lead to $\bar{N}_{\mathrm{H}_2}\propto t_{\mathrm{dur}}$, where $\bar{n}, L$, $v_{\mathrm{sh}}$ is the mean number density of a cloud, cloud size, and, shock velocity, respectively.
Therefore, we can suppose that shock duration is one of the important parameters in determining peak column density.}
Previous simulations with a high column density core/clump as an initial condition successfully demonstrated massive star/cluster formation~\citep[][]{Bonnell2004MNRAS.349..735B,krumholz2008Natur.451.1082K,Krumholz2009Sci...323..754K,Krumholz2012ApJ...754...71K}, indicating that the high column density initial core/clump leads to the massive star/cluster formation.
In \citet{Krumholz2009Sci...323..754K}, their initial condition, a dense core of 100 $M_{\odot}$ within 0.1 pc, was justified based on observed values of a protostellar object IRAS 05358+3543.
This is an extremely dense core and is not usually found.
A recent molecular study of IRAS 05358+3543 by \citet{yamada2022MNRAS.tmp.1113Y} found evidence for two colliding molecular clouds at several pc scales, which formed a dense filament including the IRAS 05358+3543 core.
This suggests that the assumed initial conditions in the previous works require advanced strong compression by an external trigger.
This suggests that the collision is an essential initial process in high mass star formation.
It is crucial that the collision duration must be longer than the {free fall time defined, using the mean density in the shock compressed layer,} time for realizing the high column density.

In this paper, we study the correlation between the shock duration and resulting star formation by controlling the amount of gas that flows into the numerical domain to examine the effect of shock duration and the origin of threshold peak column density of massive cluster formation.
{In previous related studies, there have been discrepancies in conclusions.
\citet{Inoue2018PASJ...70S..53I} and \citet{abe2021ApJ...916...83A} demonstrated that fast MHD shock compression allows for the formation of massive cores/stars by numerically setting long-lasting shock compression.
These studies assumed a long/infinite duration of a shock wave.
On the other hand, \citet{Takahira2014ApJ...792...63T} and \citet{sakre2022arXiv220507057S}, considered collisions of relativery small clouds and concluded that the fast shock compression prevent the formation of gravitationally bound cores.}
To clarify this contradictory situation, we examine the effect of shock duration on the resulting massive core/star formation.
This paper is organized as follows: In \S 2, we provide the setup of our simulations, and we demonstrated and interpreted the results in \S 3.
In \S 4, we discussed the peak column density in the shock-compressed layer by developing a simple theoretical model.
Finally, we summarize the results in \S 5.

\section{Setup for simulations} \label{sec:setup}

\begin{figure}[ht!]
\plotone{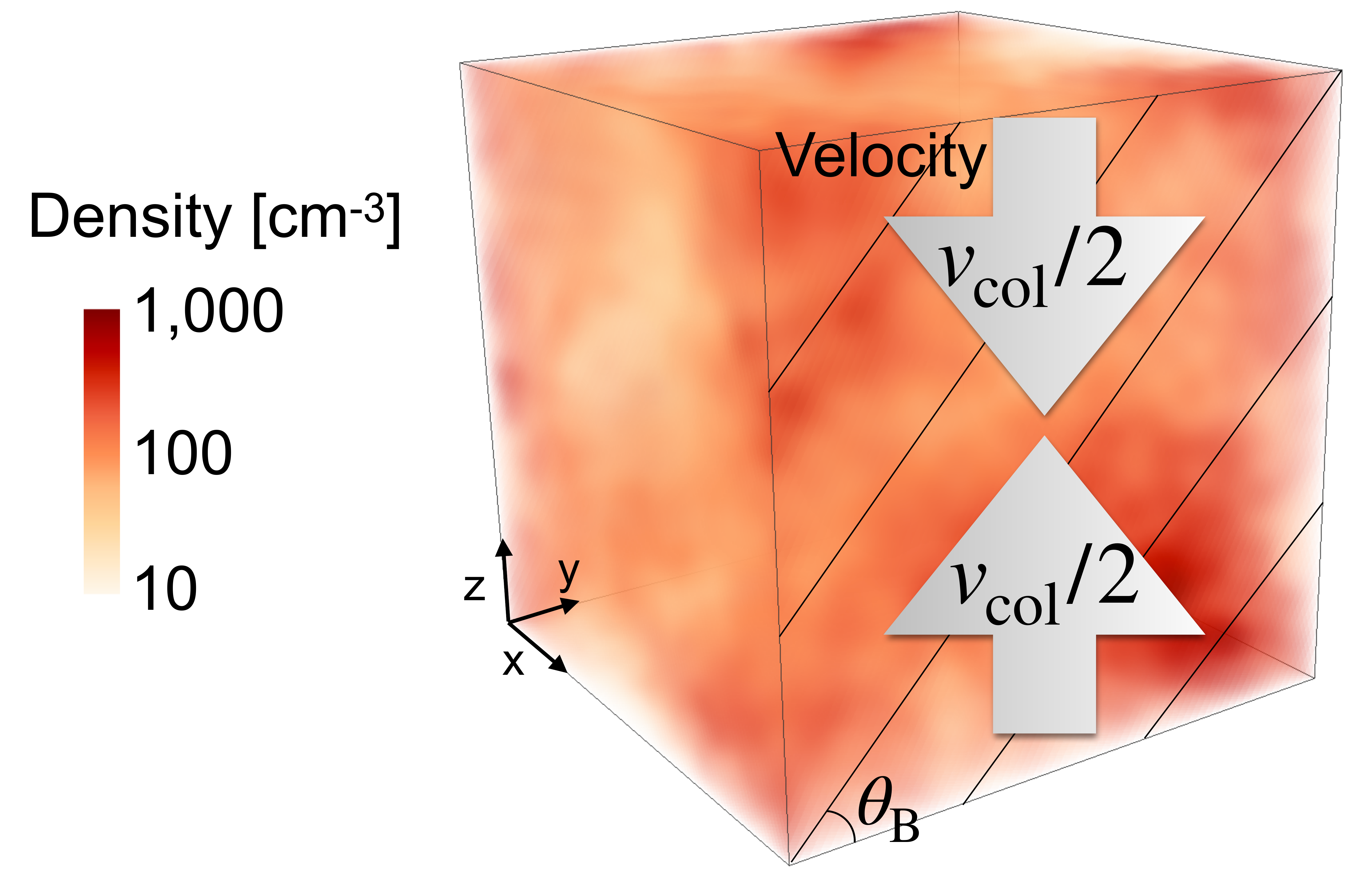}
\caption{Schematic of the initial condition.\label{fig:inicon}
The color bar represents the density magnitude, the black lines represent the initial magnetic field lines, the $\theta_{\mathrm{B}}$ represents the angle between the y-axis and the initial magnetic field lines, and the gray arrows represent the orientations of the converging flows.
}
\end{figure}

We perform three-dimensional (3D) isothermal\footnote{Isothermal treatment is a good approximation for describing compression. However, it might not be a good approximation for the expansion regions, which are created after $t_{\mathrm{dur}}$.
The minimum density in the expansion region can be as small as 1 cm$^{-3}$.
In such regions, the low-density gas would become atomic and warm.
One of the non-isothermal effects imposed on the cold gas surrounded by the warm gas is an evaporation flow due to gas heating and thermal conduction, and the flow speed is known to be subsonic \citep{inoue2006ApJ...652.1331I}.
Thus, the non-isothermal effect would not be substantial because the expansion goes on at sonic speed owing to pressure reduction.} MHD simulations including self-gravity using SFUMATO code~\citep{Matsumoto2007}.
The initial condition is the same as that of \cite{abe2021ApJ...916...83A} except that we additionally introduce a finite duration of shock compression.
In this section, the numerical setup is briefly stated.
We use a cubic numerical domain with a box size of $L_{\mathrm{box}}$ = 6 pc, which is filled with a nonuniform gas, a mean density $n_0$, and fluctuations with a power spectrum of $(\log \rho)_{\mathrm{k}}^2 \propto k^{-4}$ due to supersonic turbulence~\citep{Larson1981,Beresnyak2005,Elmegreen2004,Scalo2004,Heyer2004}.
We set the gas temperature to 10 K (corresponding sound speed $c_{\mathrm{s}} \sim$ 0.2 km/s).
We divide the numerical domain into uniform 512$^3$ cells, resulting in a spatial resolution of $\Delta x$~=~6~pc/512~=~1.2~$\times 10^{-2}$~pc.
The initial velocity field is set to, $\mathbf{v}(x,y,z) = \mathbf{v}_{\mathrm{turb}}(x,y,z) - (v_{\mathrm{col}} /2) \tanh[(z-L_{\mathrm{box}}/2)/0.1]\mathbf{\hat{z}}$, where $\mathbf{v}_{\mathrm{turb}}$ is the turbulent velocity, having a dispersion of 1.0~km~s$^{-1}$ with a power spectrum of $v_{\mathrm{k}}^2 \propto k^{-4}$, following Larson's law~\citep{Larson1981}.
This initial velocity leads to a gas collision at $z$ = $L_{\mathrm{box}}/2$ plane.
We select an initial magnetic field strength $B_0$ = 10~$\mu$G~\citep{Heiles2005a,Crutcher2012} and an angle $\theta_{\mathrm{B}}$ to the y-z plane 45$^{\circ}$, or 60$^{\circ}$.

The velocity fields at z-boundaries are given by
\begin{equation}
    \mathbf{v}^{\pm}_{\mathrm{boundary}}(x,y) = 
    \left[
    \pm {v}_{\mathrm{col}}/2 \mathbf{\hat{z}}
    + \mathbf{v}_{\mathrm{turb}}(x,y,z)
    \right]
    \exp[-(t-t_{\mathrm{stop}}) / 0.1 \mathrm{Myr}],
    \label{eqs:boundary velocity}
\end{equation}
where ``$\mathbf{v}^{+}_{\mathrm{boundary}}$" and ``$\mathbf{v}^{-}_{\mathrm{boundary}}$" are the velocity field at $z$ = 0, and 6 pc, respectively, and ``$v_{\mathrm{col}}$" is the relative velocity of two flows colliding at $z$ = 3 pc.
These boundary conditions realize the cessation of gas inflows around $t$~=~$t_{\mathrm{stop}}$, and their implementation differs significantly from that of \cite{abe2021ApJ...916...83A}.
{The density/turbulence field to be inflow from the boundary is the same density/turbulence spatial distribution as the initial condition.}
We impose free boundary conditions on the magnetic field at $z=0, L_{\mathrm{box}}$ boundaries.
We use periodic boundary conditions for all physical variables for $x=0$, $L_{\mathrm{box}}$ and $y=0$, and $L_{\mathrm{box}}$ boundary planes.

We simulate 18 different models.
Each model has a unique name, starting with ``v" (for ``velocity of collision"), followed by the collision velocity (``3," ``5," ``14," and ``18", ``24" [km s$^{-1}$]) and, the shock duration (``t"), followed by the time scale (``0.23--1.9" [Myr]).
Models with a different initial magnetic field angle and initial mean density are additionally denoted as ``ang60" and ``d300" corresponding to $\theta_{\mathrm{B}}$ = 60$^{\circ}$ and $\bar{n}_0$ = 300~cm$^{-3}$, respectively.
The set of parameters used in our simulations is listed in Table \ref{tab:modelparameters}.

In the regions where gravitational collapse occurs, we introduce the sink particle~\citep{Matsumoto2015,Inoue2018}.
The conditions for generating the sink particles depend on the resolution.
The threshold density of sink particle generation in the current simulations with 512$^3$ cells is 5.6$\times 10^{4}\mathrm{cm^{-3}}$, which is lower than that in previous studies.
Thus, we perform an AMR simulation to demonstrate whether the results depend on the resolution and threshold density of the sink creation.
The Jeans criterion is used for the refinement~\citep{Truelove1997}: $\Delta x \leq \lambda_{\mathrm{J}} / 8$ (where $\lambda_{\mathrm{J}}=\pi^{1 / 2} c_{\mathrm{s}} /\sqrt{G \rho}$ is the Jeans length), and the minimum cell size for the AMR run is $\Delta x$~=~6~pc/1024=5.9~$\times 10^{-3}$~pc.
The threshold density of sink particles is 2.2$\times 10^{5} \mathrm{cm^{-3}}$.
The results are almost the same as those in the non-AMR run, as shown in \S 3.

\begin{table*}
\caption{Model parameters.\label{tab:modelparameters}}
 \centering
  \begin{tabular}{cccccccc}
   \hline
   Model Name & $v_{\mathrm{coll}}~[\mathrm{km\ s^{-1}}]$ & $v_{\mathrm{sh}}~[\mathrm{km\ s^{-1}}]$ & $t_{\mathrm{stop}}$ [Myr] & $t_{\mathrm{dur}}$ [Myr] & $\theta_{\mathrm{B}}$ & $\bar{n}_0$ [cm$^{-3}$] & $\Delta x$ [pc]\\
   \hline \hline
    v3t1.4        & 3  & 2.2 & 0.0 & 1.4  & 45$^{\circ}$ & 100 & 1.2$\times 10^{-2}$ \\
    v5t0.94       & 5  & 3.2 & 0.0 & 0.94 & 45$^{\circ}$ & 100 & 1.2$\times 10^{-2}$ \\
    v5t1.5        & 5  & 3.2 & 0.7 & 1.5  & 45$^{\circ}$ & 100 & 1.2$\times 10^{-2}$ \\
    v14t0.39      & 14 & 7.7 & 0.0 & 0.39 & 45$^{\circ}$ & 100 & 1.2$\times 10^{-2}$ \\
    v14t0.40ang60 & 14 & 7.5 & 0.0 & 0.40 & 60$^{\circ}$ & 100 & 1.2$\times 10^{-2}$ \\
    v14t0.66      & 14 & 7.7 & 0.3 & 0.66 & 45$^{\circ}$ & 100 & 1.2$\times 10^{-2}$ \\
    v14t0.84      & 14 & 7.7 & 0.5 & 0.84 & 45$^{\circ}$ & 100 & 1.2$\times 10^{-2}$ \\
    v14t1.0       & 14 & 7.7 & 0.7 & 1.0  & 45$^{\circ}$ & 100 & 1.2$\times 10^{-2}$ \\
    v14t1.1d300   & 14 & 7.4 & 0.7 & 1.1  & 45$^{\circ}$ & 300 & 1.2$\times 10^{-2}$ \\
    v14t1.5       & 14 & 7.7 & 1.2 & 1.5  & 45$^{\circ}$ & 100 & 1.2$\times 10^{-2}$ \\
    v14t1.7ang60  & 14 & 7.5 & 1.4 & 1.7  & 60$^{\circ}$ & 100 & 1.2$\times 10^{-2}$ \\
    v14t1.9       & 14 & 7.7 & 1.7 & 1.9  & 45$^{\circ}$ & 100 & 1.2$\times 10^{-2}$ \\
    v18t0.68      & 18 & 9.7 & 0.4 & 0.68 & 45$^{\circ}$ & 100 & 1.2$\times 10^{-2}$ \\
    v24t0.23      & 24 & 13  & 0.0 & 0.23 & 45$^{\circ}$ & 100 & 1.2$\times 10^{-2}$ \\
    v24t0.43      & 24 & 13  & 0.2 & 0.43 & 45$^{\circ}$ & 100 & 1.2$\times 10^{-2}$ \\
    v24t0.61      & 24 & 13  & 0.4 & 0.61 & 45$^{\circ}$ & 100 & 1.2$\times 10^{-2}$ \\
    v24t1.6       & 24 & 13  & 1.4 & 1.6  & 45$^{\circ}$ & 100 & 1.2$\times 10^{-2}$ \\
    v24t1.6AMR    & 24 & 13  & 1.4 & 1.6  & 45$^{\circ}$ & 100 & 5.9$\times 10^{-3}$ \\
    v24t1.8       & 24 & 13  & 1.7 & 1.8  & 45$^{\circ}$ & 100 & 1.2$\times 10^{-2}$ \\
   \hline
  \end{tabular}
\end{table*}

\section{Results} \label{sec:Results}

\subsection{Column Density Maps}

\subsubsection{Short Duration Case}
\begin{figure*}
\begin{center}
\includegraphics[clip,width=15.5cm]{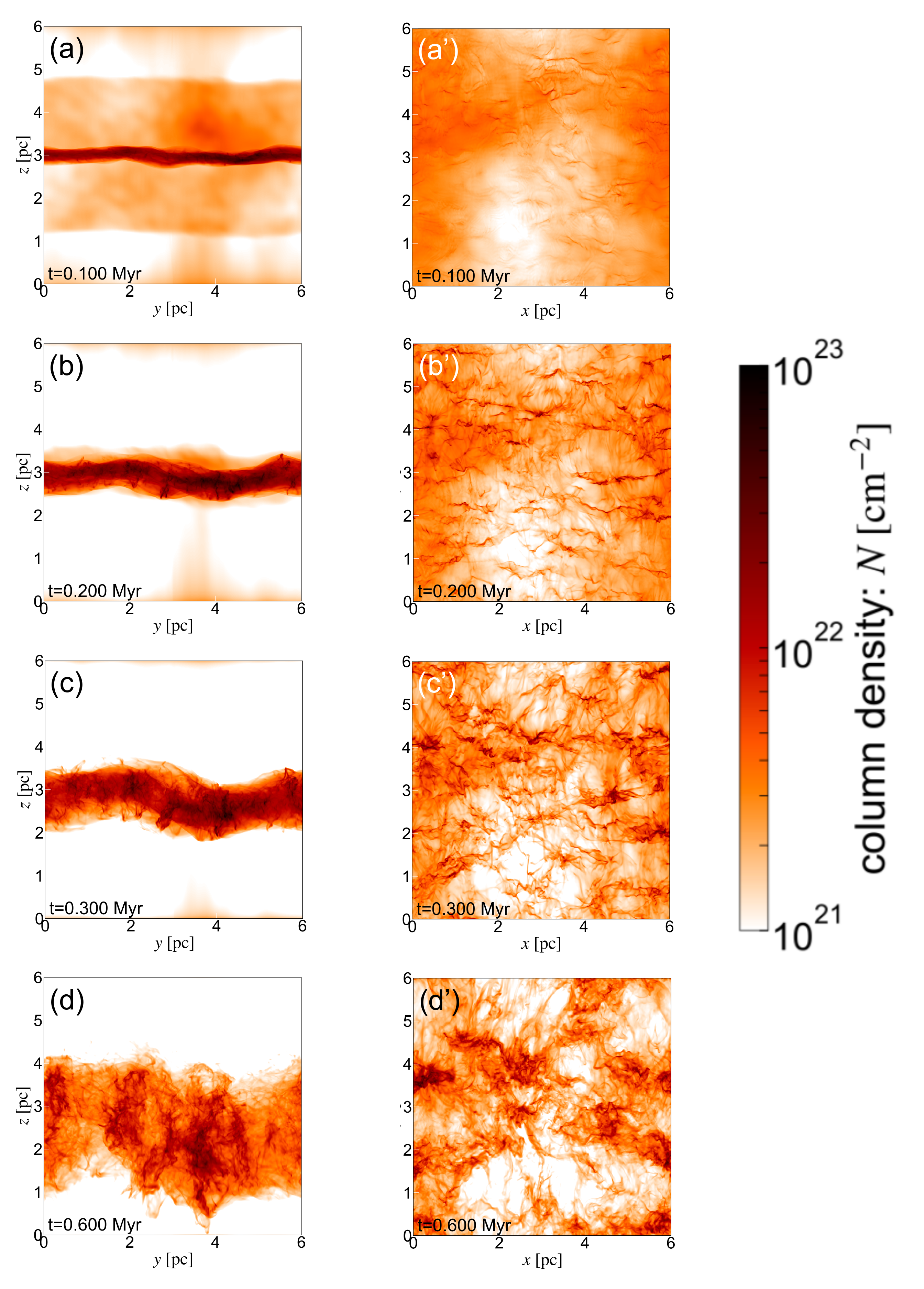}
    \caption{\small{Column density maps of the result of model v24t0.23 at time $t$ = 0.3, 0.8, 1.6, and 1.8 Myr (from top to bottom).
    \textit{Left row} (panels a, b, c, and d): y-z plane column densities.
    \textit{Right row} (panels a', b', c', and d'): x-y plane column densities.
    }}
    \label{fig:t00v24}
\end{center}
\end{figure*}
\begin{figure}
    \begin{center}
        \includegraphics[clip,width=15.5cm]{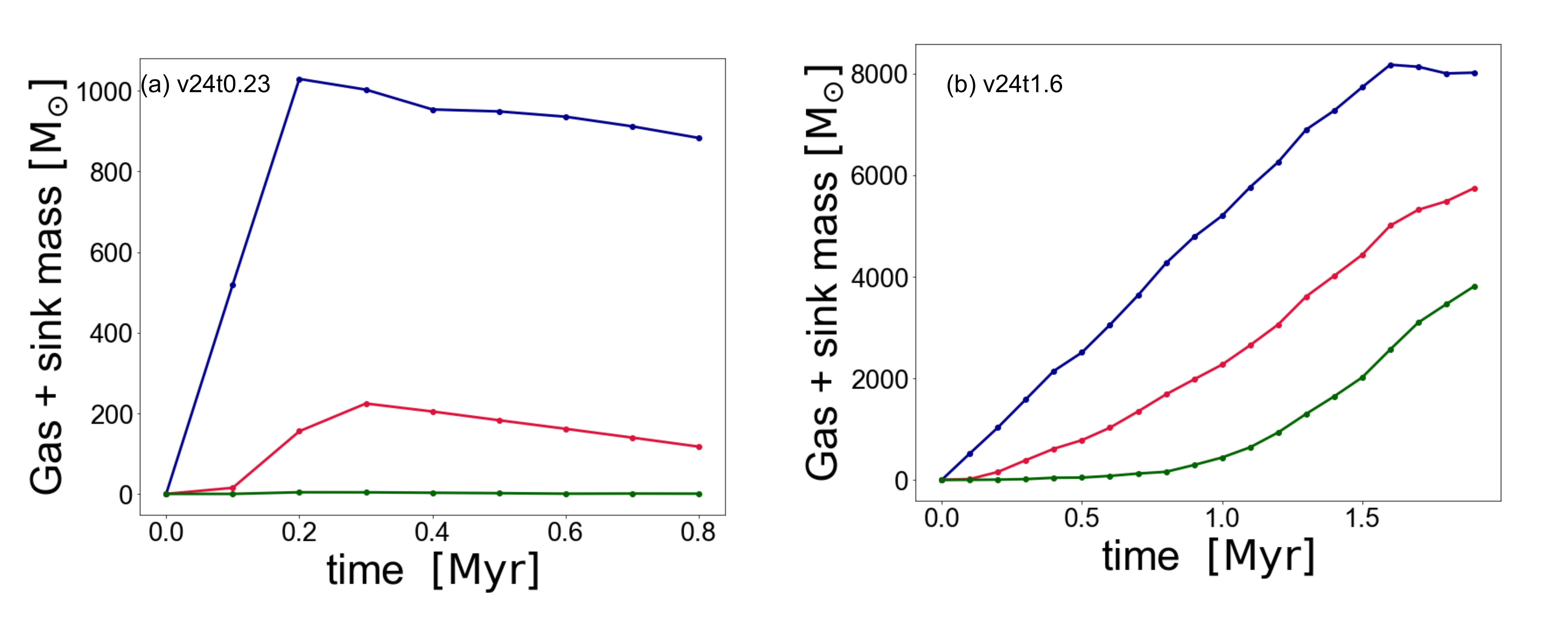}
        \caption{\small{
        Panel a: the evolution of the total dense gas mass in the model v24t0.23.
        The blue, red, and green points represent the total masses with densities above 10$^3$, 10$^4$, and 10$^5$ cm$^{-3}$, respectively.
        Panel b: the same as panel (a), but for model v24t1.6.
        }}
        \label{fig:densegasevol}
    \end{center}
\end{figure}

Figure \ref{fig:t00v24} shows snapshots of the column density map of model v24t0.23 at $t$ = 0.10, 0.20, 0.30, and 0.60 Myr.
Panels (a)--(d) and (a')--(d') show snapshots in the y-z and x-y plane slices, respectively.
The formation of dense filaments is shown in panels (a')--(d').
Recently, \citet{abe2021ApJ...916...83A} classified the filament formation mechanisms into several categories.
According to their classification, the type of filament formation seen in the results of model v24t0.23 is type O (the oblique MHD shock compression mechanism).
Note that type O filamentation works behind a strong shock of $v_{\mathrm{sh}} \gtrsim$ 5 km s$^{-1}$, and the shock waves created in this model have $v_{\mathrm{sh}}$ of approximately 13 km s$^{-1}$.
The evolution of the mass of the dense gas in the numerical domain is shown in panel (a) of Figure \ref{fig:densegasevol}.
The blue, red, and green points represent the gas masses of the regions with densities greater than 10$^3$, 10$^4$, and 10$^5$ cm$^{-3}$, respectively.
Panels (b')--(d') in Figure \ref{fig:t00v24} and panel (a) in Figure \ref{fig:densegasevol} show that the mass of the dense gas decreases 0.2--0.3 Myr.
This evolution is due to the expansion of the shock-compressed layer and the resulting pressure reduction {(see Figure \ref{fig:t00v24}d)}, which deconfines dense filaments.
The time at which the compression layer starts to expand $t_{\mathrm{dur}}$ can be written using $t_{\mathrm{stop}}$ as
\begin{eqnarray}
    t_{\mathrm{dur}} &\simeq& t_{\mathrm{stop}} + \frac{L_{\mathrm{box}}/2 - {v_{\mathrm{1}}} t_{\mathrm{dur}} }{ {v}_{\mathrm{col}}/2} \label{eqs:t duration1} \\
    &\simeq& \frac{t_{\mathrm{stop}} + L_{\mathrm{box}}/{v}_{\mathrm{col}}}{1 + \sqrt{2} v_{\mathrm{alf,0}} / v_{\mathrm{col}}}.
   \label{eqs:t duration2}
\end{eqnarray}
where $v_1 \simeq \bar{v}_{\mathrm{Alf,\perp}} / \sqrt{2}$ is the shock velocity at the rest frame of the compression layer, and $\bar{v}_{\mathrm{Alf,\perp}} = B_{0,\perp}/\sqrt{4\pi\bar{\rho}_0}$ is the mean Alfv\'{e}n velocity.
$B_{0,\perp} = B_{0} \cos(\theta_{\mathrm{B}})$ is the initial magnetic field strength perpendicular to the shock normal.
The second term on the right-hand side of Eq. (\ref{eqs:t duration1}) represents a retarded time for the gas to reach the compression layer after $t_{\mathrm{stop}}$.
In the case of model v24t0.23, $t_{\mathrm{dur}} \simeq 0.23$ Myr corresponds to the time when the dense gas mass in Figure \ref{fig:densegasevol} reaches its peak.

The free-fall time in the postshock layer, which gives the timescale for self-gravitating sheet fragmentation~\citep[][]{Nagai1998}, can be estimated as
\begin{eqnarray}
    t_{\mathrm{ff}}
&=&
\sqrt{\frac{1}{2\pi G\bar{\rho}_1}}\nonumber \\
&=&
\sqrt{\frac{\bar{v}_{\mathrm{Alf,\perp}}}{2 \sqrt{2}\pi G \bar{\rho}_0 \bar{v}_{\mathrm{sh}}}}\nonumber \\
&=&
\sqrt{\frac{B_{0,\perp}}{4\sqrt{2}\pi^{3/2} G\bar{\rho}^{3/2}_{0} \left( v_{\mathrm{coll}}/2 + B_0/\sqrt{8\pi \bar{\rho}_0} \right)}}\nonumber \\
&\simeq &
 1.0\ \mathrm{Myr}
 \left(
\frac{B_{0,\perp}}{10\ \mathrm{\mu G}}
\right)^{1/2}
\left(
\frac{\bar{n}_{0}}{100\ \mathrm{cm^{-3}}}
\right)^{-3/4}\nonumber\\
&\times & 
\left[ 
\left(
\frac{v_{\mathrm{coll}}}{12\ \mathrm{km/s}}
\right)
+ 0.17
\left(
\frac{B_{0,\perp}}{10\ \mathrm{\mu G}}
\right)
\left(
\frac{\bar{n}_{0}}{100\ \mathrm{cm^{-3}}}
\right)^{-1/2}
\right]^{-1/2},
\label{equation:freefalltime}
\end{eqnarray}
where $\bar{\rho}_1 \simeq \sqrt{2}\mathcal{M}_{\mathrm{A}} \bar{\rho}_{\mathrm{0}}$ is the mean density of the shocked layer \citep[e.g.,][]{inoue2013ApJ...774L..31I}, and $\bar{v}_{\mathrm{sh}} = v_{\mathrm{coll}}/2+v_1 \simeq v_{\mathrm{coll}}/2 + B_0/\sqrt{8\pi \bar{\rho}_0}$ represents the mean shock velocity.
Because {$t_{\mathrm{dur}} < t_{\mathrm{ff}} \simeq 0.68$ Myr} in model v24t0.23, the shock compression layer expands before gravitational fragmentation and no sink particles are created.


\subsubsection{Long Duration Case}
\begin{figure*}
\begin{center}
\includegraphics[clip,width=15.5cm]{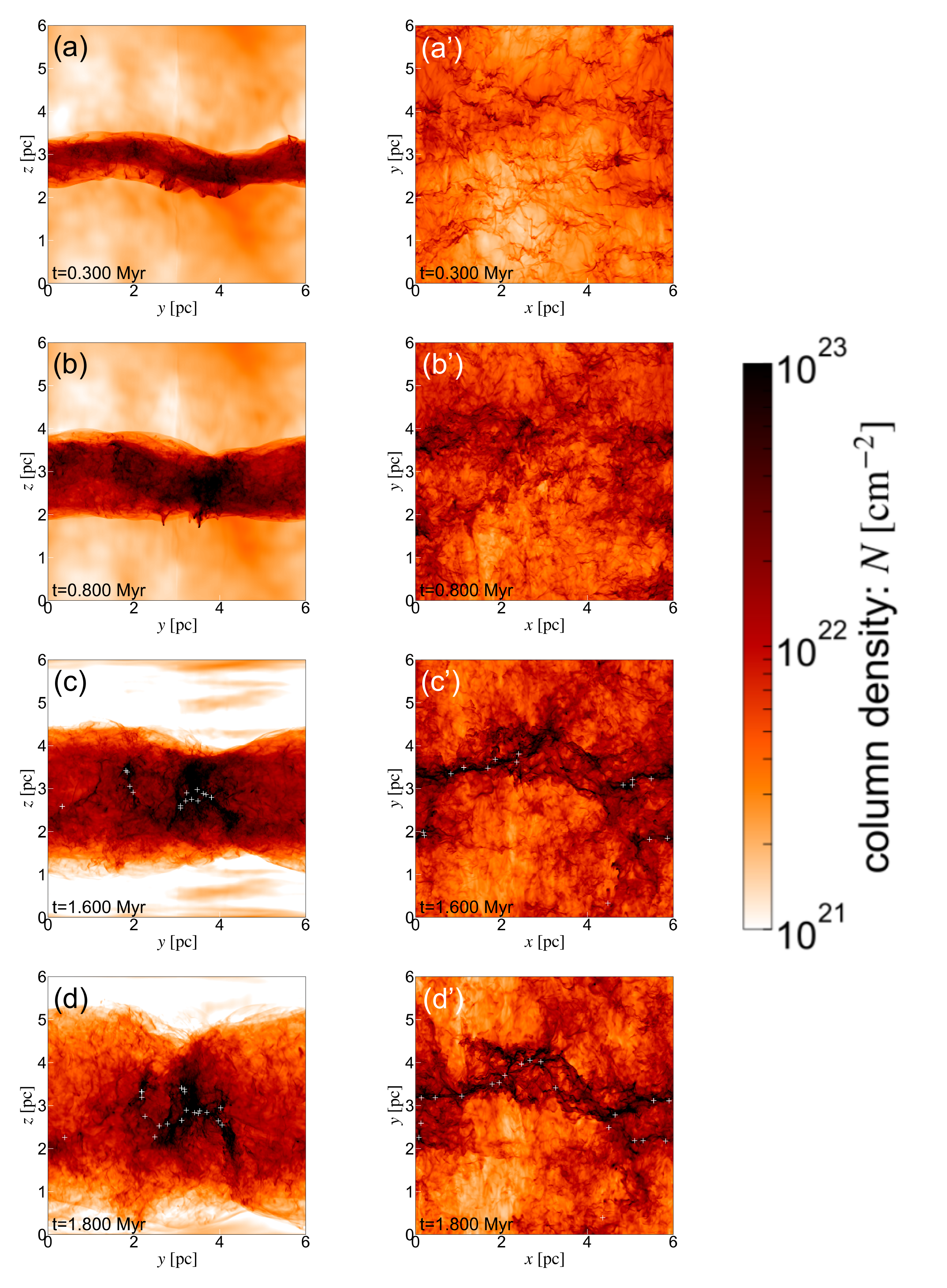}
    \caption{\small{Column density maps in the result of model v24t1.6 at time $t$ = 0.3, 0.8, 1.6, and 1.8 Myr (from top to bottom).
    \textit{Left row} (panels a, b, c, and d): y-z plane column densities.
    \textit{Right row} (panels a', b', c', and d'): x-y plane column densities.
    }}
    \label{fig:t14v24}
\end{center}
\end{figure*}
Similar to Figure \ref{fig:t00v24}, Figure \ref{fig:t14v24} shows snapshots of the column density map of model v24t1.6 at t = 0.30, 0.80, 1.60, and 1.80 Myr.
Because the shock velocity is the same as in the previous model (v24t0.23), dense filamentary structures are created via the type O mechanism.
The ``+'' symbol{s} in Figure \ref{fig:t14v24} indicate the positions of the sink particles.
The evolution of the dense gas mass in the numerical domain is shown in panel (b) of Figure \ref{fig:densegasevol}.
The blue, red, and green points represent the masses of the regions with densities greater than 10$^3$, 10$^4$, and 10$^5$ cm$^{-3}$, respectively, as in panel (a).
{It should be noted that the range of the vertical axis is different from that of the panel (a).}
{In this model also}, the shock-compressed layer expands after $t = t_{\mathrm{dur}}\simeq 1.6$ Myr, as in model v24t0.23.
However, the duration of the compression is longer than the {free-fall-time calculated using the mean density in the layer} $t_{\mathrm{ff}} \simeq$ 0.58 Myr {(Eq. \ref{equation:freefalltime})}, allowing the filaments to coalesce because of the gravitational contraction of the compression layer.
Therefore, in contrast to model v24t0.23, the dense gas mass continues to increase even after $t_{\mathrm{dur}}$ (panel a in Figure \ref{fig:densegasevol}).

\subsection{Peak Column Density vs. Dense Gas Mass} \label{subsec:PeakCDvsDenseGasMass}
Observations show that the peak column density of a star-forming cloud correlates with the number of OB stars in the system~\citep[][]{enokiya2021PASJ...73S..75E}.
Here, we compare the results of our simulations with the observed results.
\begin{figure}
    \begin{center}
        \includegraphics[clip,width=15.5cm]{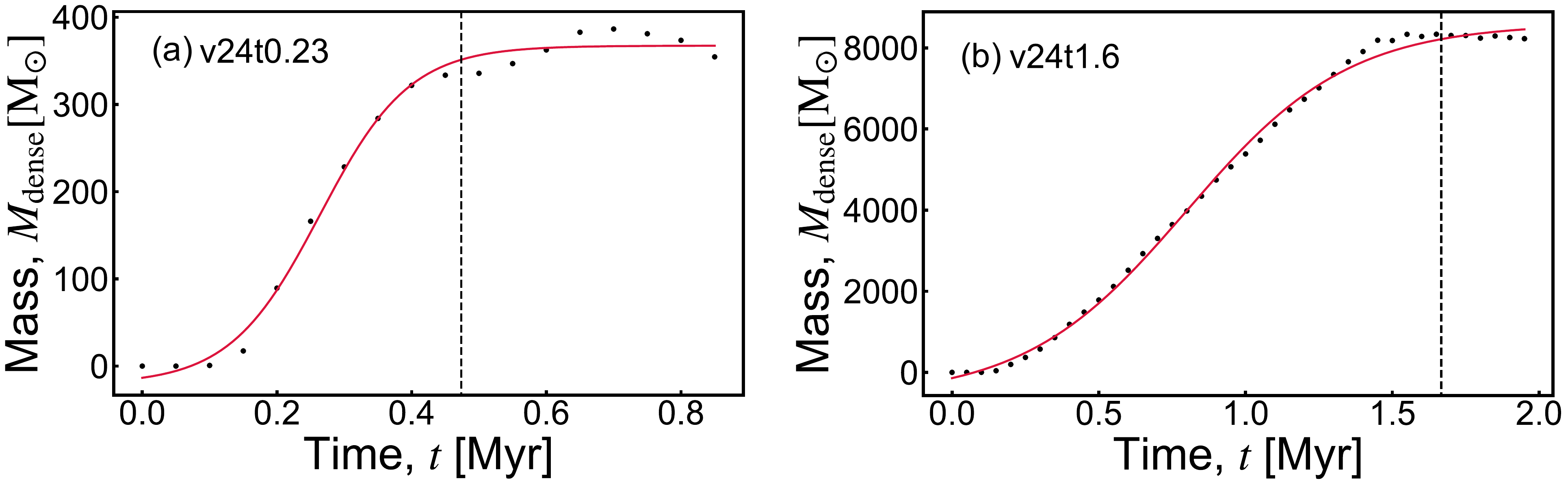}
        \caption{\small{
         Evolution of $M_{\mathrm{dense}}(t) = M_{\mathrm{sink,tot}}(t) + M_{\mathrm{Av8}}(t)$ for the results of models v24t0.23 (left) and v24t1.6 (right), and the fitting curve obtained using a trial function $\alpha \tanh[2\pi (t-t_0)/t_{\mathrm{w}}] + \beta$ (red line), where $\alpha, t_0, t_{\mathrm{w}}, \beta$ are fitting parameters. 
        Time $t_{\mathrm{sat}} \equiv t_0 + t_{\mathrm{w}} /4$ is defined as the time at which the total gas mass associated with the star formation is measured (dashed line).
        }}
        \label{fig:av8evol}
    \end{center}
\end{figure}
\begin{figure}
    \begin{center}
        \includegraphics[clip,width=15.5cm]{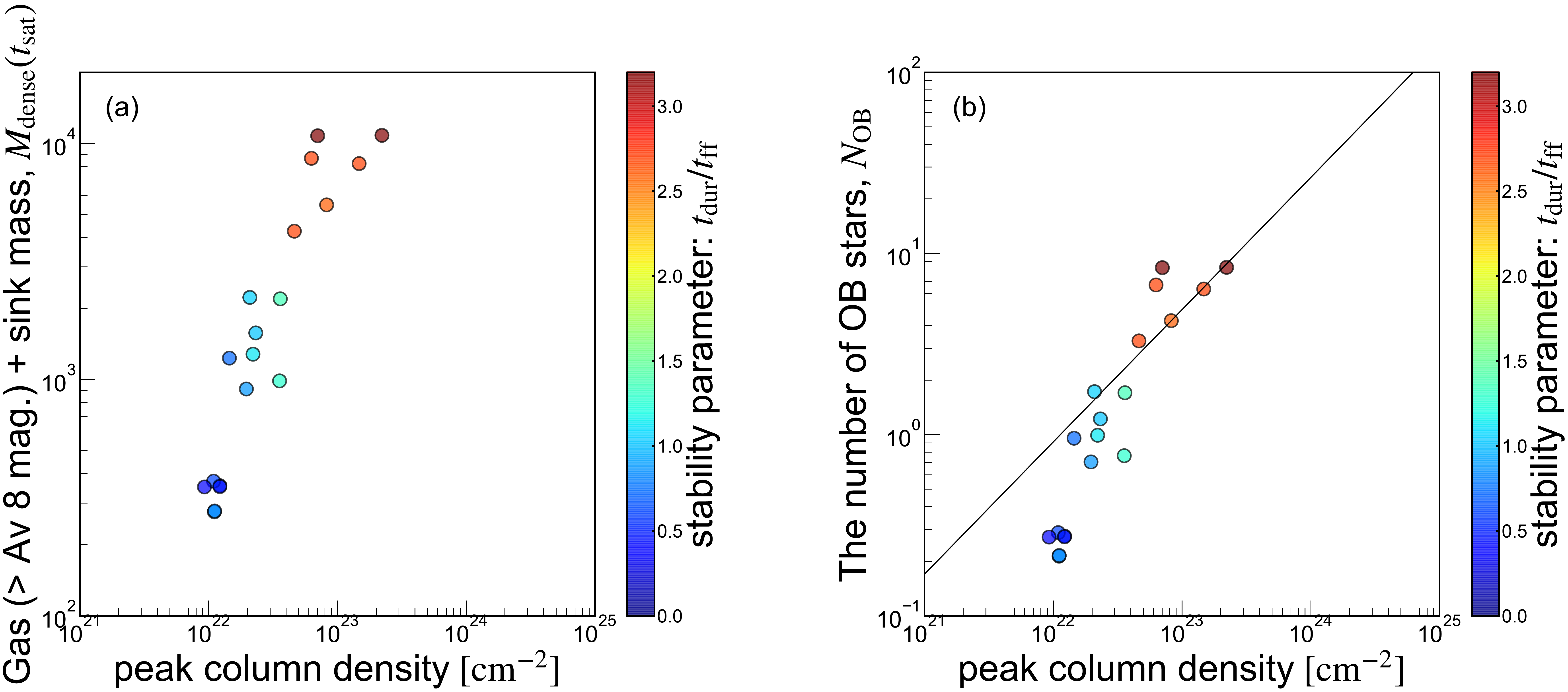}
        \caption{\small{
        Panel a: Scatter plot of the peak column density vs. the dense gas mass $M_{\mathrm{dense}}(t_{\mathrm{sat}})$ for all runs summarized in Table \ref{tab:modelparameters}.
        Before computing the peak column density of the system, we take a smoothing of the column density using the Gaussian kernel function of a width of 0.5 pc, corresponding to the typical beam width for massive star-forming regions.
        The color of the points represents $t_{\mathrm{dur}}/t_{\mathrm{ff}}$ indicating the influence of self-gravity.
        Panel b: Scatter plot of the peak column density vs. the estimated number of OB-type stars $N_{\mathrm{OB}}$.
        $N_{\mathrm{OB}}$ is estimated using Eq. (\ref{eqs:the number of ob star}).
        {The black line is the fitting line in Figure 9 (b) of \citet{enokiya2021PASJ...73S..75E}.}
        }}
        \label{fig:enkyDiagram_repr_abe}
    \end{center}
\end{figure}
We consider the following two types of mass in the numerical domain, as a proxy of star formation: the total mass of sink particle, $M_{\mathrm{sink,tot}}(t)$, and total gas mass in the region with $A_V$ $>$ 8 mag. ({$N_{\mathrm{H}_2} \simeq$} 7.8$\times$10$^{21}$ cm$^{-2}$), $M_{\mathrm{Av8}}(t)$~\citep{lada2010ApJ...724..687L}.
The sum of these masses $M_{\mathrm{dense}}(t) \equiv M_{\mathrm{sink,tot}}(t) + M_{\mathrm{Av8}}(t)$, {is} proportional to the total mass of stars in the system.
Figure \ref{fig:av8evol} shows the evolution of $M_{\mathrm{dense}}(t)$.
Panels (a) and (b) show the results for models v24t0.23 and v24t1.6, respectively.
{In our simulation,} the $M_{\mathrm{dense}}(t)$ saturates in late time, and we can define a mass associated with star formation.
By fitting the $M_{\mathrm{dense}}(t)$ using a trial function $\alpha \tanh[2\pi (t-t_0)/t_{\mathrm{w}}] + \beta$, as shown in the red lines in Figure \ref{fig:av8evol}, we obtain the time $t_{\mathrm{sat}} \equiv t_0 + t_{\mathrm{w}} /4$ (dashed lines in Figure \ref{fig:av8evol}) at which we measure total dense gas mass representing the star formation activity in the system, where $\alpha, t_0, t_{\mathrm{w}}, \beta$ are fitting parameters.
Note that $t_{\mathrm{sat}}$ corresponds to the time when the fitting curve reaches {approximately $\alpha\tanh[2\pi (t_{\mathrm{sat}}-t_0)/t_{\mathrm{w}}]+ \beta = \tanh(\pi/2)+ \beta \simeq 0.92\alpha+ \beta$, i.e., this value means mass corresponding to (0.92+1.0)$\times$ 100/2 = 96 [\%] of the increment in the trial function.}

Generally, a peak column density highly depends on a spatial grid (or in other words, spatial resolution) used for the derivation.
Thus, the observed peak column densities of cloud-cloud collision (CCC) candidates in the plots of Figure 9 in \citet{enokiya2021PASJ...73S..75E} have potentially independent spatial resolutions. Nevertheless, the correlation between the peak column density and the number of O- and B-type stars is apparent in the plots.
This suggests that the spatial resolutions are similar among the observed CCC candidates. We have checked the spatial resolution among the CCC candidates and found it roughly ~ the size of the smaller cloud/10.
Assuming 5 pc for the size of the smaller cloud \citep[see Figure 9 in][]{Fukui2021}, we estimate the typical spatial resolution of the observed CCC candidates to be 0.5 pc.
Thus, before computing the peak column density of the system, we take a smoothing of the column density structure using the Gaussian kernel function of a width of 0.5 pc.
In panel (a) of Figure \ref{fig:enkyDiagram_repr_abe}, we demonstrated the scatter plot of the peak column density vs. the dense gas mass $M_{\mathrm{dense}}(t_{\mathrm{sat}})$ for all runs listed in Table \ref{tab:modelparameters}.
The color of the points indicates $t_{\mathrm{dur}}/t_{\mathrm{ff}}$, which represents the influence of self-gravity.
{We employ the following procedures to estimate the number of OB stars formed in the numerical system:
We multiply $M_{\mathrm{dense}}(t_{\mathrm{sat}})$ by the star formation efficiency $\mathrm{SFE}_{\mathrm{dense}}\sim$0.1 to obtain the total mass of stars~\citep{Fukui2021}.
The fraction of OB-type stars in the total stellar mass $f_{\mathrm{OB}}=\left(\int_{10M_{\odot}}^{150M_{\odot}}dM\ M F_{\mathrm{Kroupa}}\right) / \left(\int_{0.01M_{\odot}}^{150M_{\odot}}dM\ M F_{\mathrm{Kroupa}} \right) = 0.19$ is then multiplied to obtain the total mass of OB stars expected in the system.
Note that we consider stars heavier than the early-B type, i.e., larger than approximately 10~$M_{\mathrm{\odot}}$.
Here, $F_{\mathrm{Kroupa}} \equiv dN/dM$ is an IMF proposed by \citet[][]{kroupa2001MNRAS.322..231K}.
Therefore, the estimation of the number of OB stars is
\begin{eqnarray}
    N_{\mathrm{OB}} 
    &\simeq& 
    M_{\mathrm{dense}}(t_{\mathrm{sat}}) \times \mathrm{SFE}_{\mathrm{dense}} \times f_{\mathrm{OB}}
    \times \left(
    \frac{\int_{10M_{\odot}}^{150M_{\odot}}dM\ M F_{\mathrm{Kroupa}}}{\int_{10M_{\odot}}^{150M_{\odot}}dM\ F_{\mathrm{Kroupa}}}
    \right)^{-1} ,
   \label{eqs:the number of ob star}
\end{eqnarray}
where $\int_{10M_{\odot}}^{150M_{\odot}}dM\ M F_{\mathrm{Kroupa}}/\int_{10M_{\odot}}^{150M_{\odot}}dM\ F_{\mathrm{Kroupa}}$ on the right-hand side is the typical mass of OB stars.}
Panel (b) in Figure \ref{fig:enkyDiagram_repr_abe} is compiled from panel (a) using Eq. (\ref{eqs:the number of ob star}).
The result shows a correlation similar to that of~\citet[][]{enokiya2021PASJ...73S..75E}, although our simulations cover only a limited range of peak column density than the observations.
The power-law index obtained by fitting data in panel (b) is 0.75, which agrees well with that of observations 0.73$\pm 0.11$~\citep[][]{enokiya2021PASJ...73S..75E}.
The black line is the fitting line in Figure 9 (b) of \citet{enokiya2021PASJ...73S..75E}.
It has been known observationally that massive star cluster formation occurs in the clouds with a peak column density greater than $N_{\mathrm{peak}} \geq 10^{23}\ \mathrm{cm}^{-2}$~\citep[][]{fukui2016ApJ...820...26F}.
Our simulations agree with this threshold peak column density {(see the region with $N_{\mathrm{peak}} \sim 10^{23} \mathrm{cm^{-2}}$ in panel (b) of Figure \ref{fig:enkyDiagram_repr_abe})}.
The stability parameter, ratio of the shock duration and the free-fall time of the compressed layer $t_{\mathrm{dur}}/t_{\mathrm{ff}}$, indicates that massive stars are expected to be formed where the dense shock-compressed layer is kept over the free-fall time.
{We find that when the $N_{\mathrm{peak}}$ exceeds $10^{23}\ \mathrm{cm}^{-2}$ the stability parameter $\gtrsim$ 2.}
Coincidentally, the threshold peak column density takes a similar value to the threshold column density for fragmentation suppression $1\ \mathrm{g\ cm}^{-2}$ proposed by~\citet[][]{krumholz2008Natur.451.1082K}.

\subsection{Sink Mass Histogram}
\begin{figure}[ht!]
\plotone{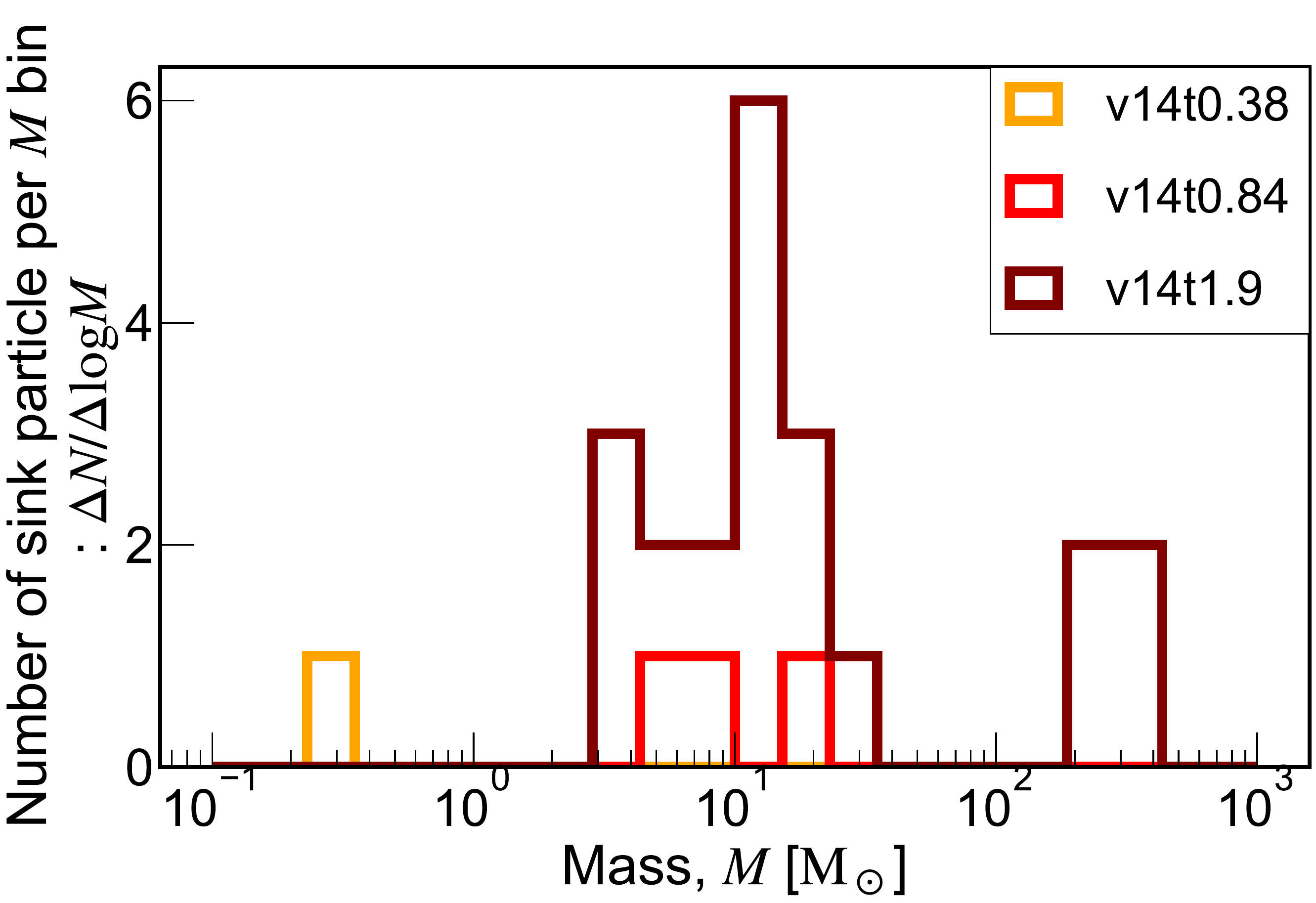}
\caption{Sink mass histogram at $t = t_{\mathrm{sat}}$ in the results of models v14t0.38 (orange), v14t0.84 (red), and v141.9 (brown).\label{fig:sinkhistevol}
}
\end{figure}
\begin{figure}[ht!]
\plotone{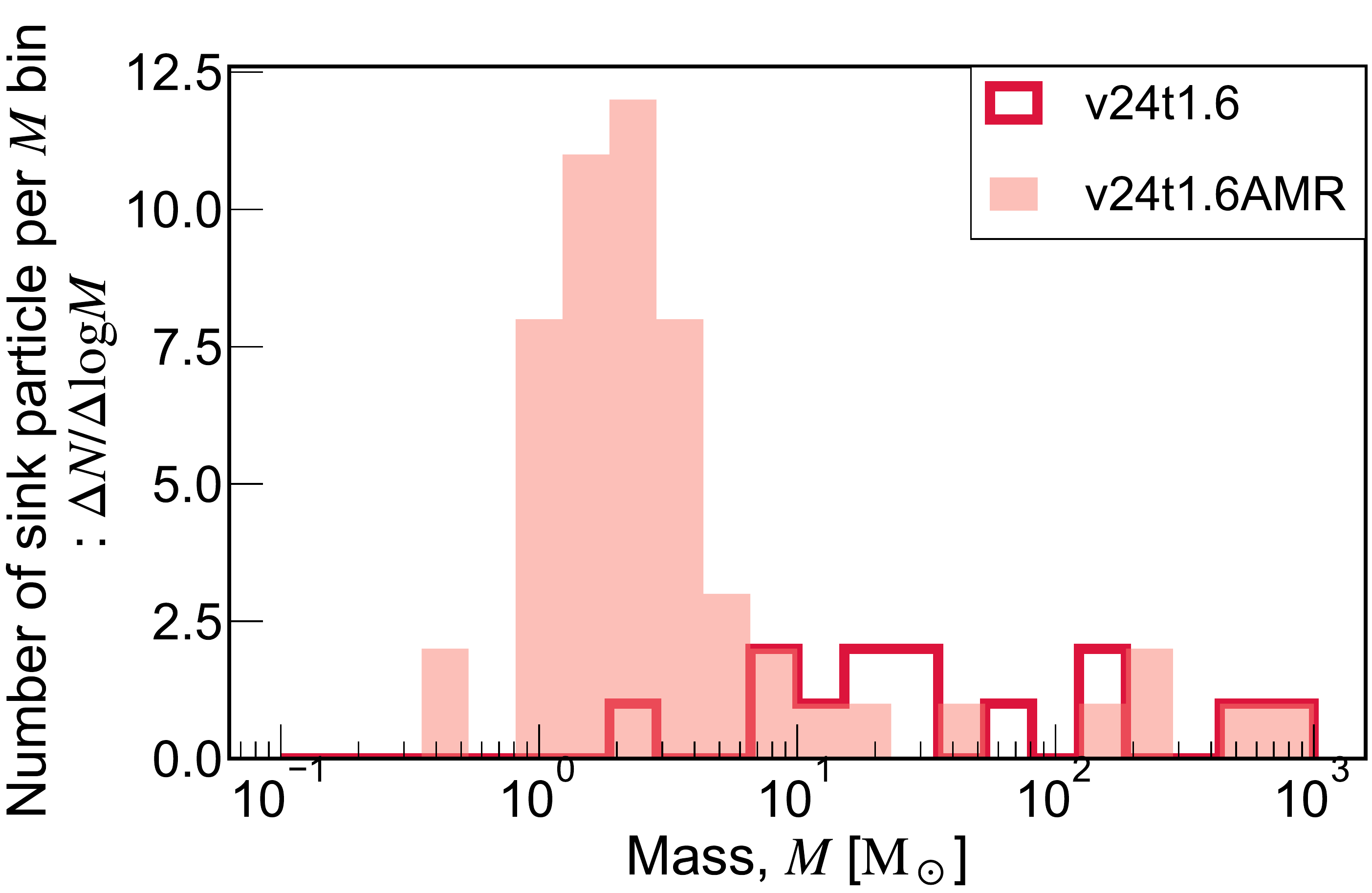}
\caption{Sink mass histogram at $t = t_{\mathrm{sat}}$ in the models v24t1.6 (red) and v24t1.6AMR. (filled-in region)\label{fig:sinkhistAMR}
}
\end{figure}
Figure \ref{fig:sinkhistevol} shows the sink mass histograms at $t = t_{\mathrm{sat}}$ for the models v14t0.38, v14t0.84, and v14t1.9.
In our simulations, the sink particles correspond to gravitationally collapsing cores.
Note that Figure \ref{fig:sinkhistevol} is not a core mass function because low mass cores are not well resolved in our simulations.
a massive sink particle of 355 $M_{\odot}$, and OB-type stars can be formed in the model v14t1.9, which has a long flow duration of $t_{\mathrm{dur}}/t_{\mathrm{ff}} \simeq 2.6$.
The mass/number of sink particles increases with $t_{\mathrm{dur}}$, as expected, but only the long duration model v14t1.9 has $t_{\mathrm{dur}}/t_{\mathrm{ff}}\gtrsim 2$ and exhibits active massive star formation.

Figure \ref{fig:sinkhistAMR} shows a sink mass histogram at $t = t_{\mathrm{sat}}$ in the models v24t1.6 and v24t1.6AMR, comparing the results with and without AMR.
The number of sink particles with masses $\sim\ 1\ M_{\odot}$ increased because AMR calculation can capture self-gravitational fragmentation of short-wavelength modes in dense filaments.
However, the total masses of massive sink particles ($> 10 M_{\odot}$) are almost the same between v24t1.6 ($M_{\mathrm{sink}}=1803\ M_{\odot}$) and v24t1.6AMR ($M_{\mathrm{sink}}=2048\ M_{\odot}$), indicating that the results exhibited in \S 3.2 do not change significantly, even if the spatial resolution is improved.



\section{Discussion}
In \S \ref{subsec:PeakCDvsDenseGasMass}, we stated that self-gravitational collapse must achieve a peak column density of 10$^{23}$~cm$^{-2}$.
Given that the column density map obtained from observations approximately $\sim$ 500 $\times$ 500 pixels, the peak column density approximately corresponds to 5$\sigma$ away from the mean of the PDF obtained from observations.
Here, we estimate the peak column density using a theory of the gas density PDF in a turbulent medium, which shows that gravitational collapse in shock-compressed layers is required to achieve a peak column density of 10$^{23}$ cm$^{-2}$.
The PDF of the gas density in a turbulent medium shows log-normal PDF of the form~\citep[][]{passot2003A&A...398..845P,Padoan2014prpl.conf...77P}
\begin{equation}
p_{\mathrm{s}}(s)=\frac{1}{\sqrt{2 \pi \sigma_{s}^{2}}} \exp \left(-\frac{\left(s-s_{1}\right)^{2}}{2 \sigma_{\mathrm{s}}^{2}}\right).
\end{equation}
{where $s \equiv \ln \left(\rho / \rho_{1}\right)$, $\rho_{\mathrm{1}}$ and $s_{\mathrm{1}}$ represent the mean density and mean logarithmic density in the shocked layer, respectively.
The latter is related to the standard deviation $\sigma_{\mathrm{\mathrm{s}}}$ as $s_{1}=-\sigma_{\mathrm{s}}^{2}/2$.}
{\citet{Molina2012MNRAS.423.2680M} find that} the standard deviation $\sigma_{\mathrm{s}}$ is determined using the turbulence Mach number $\mathcal{M}$, plasma beta $\beta$, and the ratio of solenoidal mode to compressive mode $b$ (pure solenoidal forcing gives $b$=1/3 and pure compressive forcing gives $b$=1) as
\begin{equation}
\sigma_{\mathrm{s}}^{2}=\ln \left(1+b^{2} \mathcal{M}^{2} \frac{\beta}{\beta+1}\right).
\end{equation}
Here, we estimate the maximum possible column density of a cloud formed by the gas collision.
The density PDF can have maximum width if we substitute the collision velocity for the turbulence velocity dispersion
\begin{equation}
\mathcal{M}_{\mathrm{max}}\equiv v_{\mathrm{col}}/(2c_{\mathrm{s}}).
\end{equation}
The beta of the shock-compressed layer is given by the shock jump condition of the isothermal MHD as
\begin{equation}
\beta_{\mathrm{sh}} = \frac{8 \pi \rho_{\mathrm{1}} c_s^2}{B_{\mathrm{1}}^2} \simeq \frac{8 \pi \rho_{\mathrm{0}} c_s^2}{\sqrt{2} \mathcal{M}_{\mathrm{A,max}} B_{\mathrm{0}}^2},
\end{equation}
where $\rho_{\mathrm{1}}\simeq \sqrt{2} \mathcal{M}_{\mathrm{A,max}}\rho_{\mathrm{0}}$, $B_{\mathrm{1}}\simeq \sqrt{2} \mathcal{M}_{\mathrm{A,max}}B_{\mathrm{0}}$, and $\mathcal{M}_{\mathrm{A,max}}=v_{\mathrm{col}}/(2v_{\mathrm{alf,0}})$ are the density, magnetic field in the compression layer, and the Alfv\'{e}n Mach number of the shock.
If we use the parameter $b$=0.4, which is expected in the shock-compressed layer~\citep[][]{kobayashi2022arXiv220300699K}, then we can determine $\sigma_{\mathrm{s,max}}$ and density PDF.

Numerical experiments of supersonic isothermal turbulence and observations~\citep{Goodman2009ApJ...692...91G} show that the PDF of column density is also close to log-normal~\citep{Federrath2010A&A...512A..81F}
\begin{equation}
p_{\mathrm{\eta}}(\eta)=\frac{1}{\sqrt{2 \pi \sigma_{\eta}^{2}}} \exp \left(-\frac{\left(\eta-\eta_{1}\right)^{2}}{2 \sigma_{\mathrm{\eta}}^{2}}\right),
\end{equation}
where $\eta \equiv \ln \left(N / N_{1}\right)$.
{$N$ and $N_1$ are column density and the mean column density, respectively.}
By comparing $s_{\mathrm{1}}$ and $\eta_{\mathrm{1}}$ in numerical simulations by \citet[][]{Federrath2010A&A...512A..81F}, $\eta_{\mathrm{1}}$ is approximately expressed as $\eta_{\mathrm{1}} \sim 0.5 s_{\mathrm{1}}$.
The standard deviation is also approximated by $\sigma_{\mathrm{\eta}} \sim 0.5 \sigma_{\mathrm{s}}$.
These relationships can be used to convert density PDF to column density PDF.

The peak column density is defined as 5$\sigma$ away from the mean
\begin{equation}
N_{\mathrm{5\sigma}} 
\equiv 
\exp(\eta_1 + 5 \sigma_{\mathrm{\eta,max}}) N_1,
\end{equation}
where $\sigma_{\mathrm{\eta,max}} \equiv 0.5 \sigma_{\mathrm{max}}$.
By substituting the parameters of the strong collision case ($v_{\mathrm{col}} = 10$ km s$^{-1}$, $\bar{n}_0 = 100$ cm$^{-3}$, and $B_0 = 10\ \mathrm{\mu G}$), we obtain the peak column density $N_{\mathrm{5\sigma}}$ = 8.7$\times$10$^{22}$ cm$^{-2}\ <$ 10$^{23}$ cm$^{-2}$.
Therefore, even if we use an overestimated velocity dispersion of the turbulence in the shock-compressed layer, the estimated peak column density of the structure created only by turbulence does not exceed 10$^{23}$ cm$^{-2}$, indicating that gravitational collapse of the shock layer must achieve 10$^{23}$ cm$^{-2}$.
In other words, the shock-compressed layer must be maintained until the gravitational collapse to form a massive star cluster, which enhances the peak column density, starts to activate.


Recently \citet[][]{sakre2022arXiv220507057S} have pointed out that, for a given initial cloud, there is a maximum collision speed for {triggering} of star formation by a cloud collision.
This is consistent with our results because high collision speed leads to a short shock duration.
However, this does not mean that high shock velocity is negative for induced star formation.
As we discuss below, the high shock velocity is generally positive for star formation:
The duration of the shock is estimated by $t_{\rm dur}=L/v_{\rm sh}$, where $L$ is the spatial scale and $v_{\rm sh}$ is the typical velocity of the flows.
If we suppose supersonic turbulent gas flow collision as an origin of the shock, the Larson's law gives $v_{\rm sh}\propto L^{0.5}$ or $L\propto v_{\rm sh}^2$, and then, $t_{\rm dur}=L/v_{\rm sh}\propto v_{\rm sh}$.
Since the average free-fall time of the shocked layer is $t_{\rm ff}\propto v_{\rm sh}^{-1/2}$ from Eq.~(\ref{equation:freefalltime}), 
the ratio of the shock duration and the postshock free-fall time is written as $t_{\rm dur}/t_{\rm ff}\propto v_{\rm sh}^{3/2}$.
This indicates that faster shock leads to a longer duration in units of the free-fall time, i.e., larger clouds naturally lead to more active star formation.
In contrast, if we fix the scale of the flow $L$ and use a different scaling $t_{\rm dur}=L/v_{\rm sh}\propto v_{\rm sh}^{-1}$, we obtain an opposite result $t_{\rm dur}/t_{\rm ff}\propto v_{\rm sh}^{-1/2}$.
Thus, under the fixed cloud-scale, faster shock leads to a negative effect on star formation, but we should bear in mind that this is due to artificially fixed $L$ and not a general trend.

\section{Summary}
We performed the shock compression simulations of molecular clouds using 3D isothermal MHD code with self-gravity (SFUMATO).
The effect of the duration of the shock-compressed layer on filament and star formation was investigated by treating shock duration as a controlling parameter.
We examined the relationship between peak column density and the estimated number of OB-type stars to understand the initial conditions of massive star formation and compared it with the observation by \citet[][]{enokiya2021PASJ...73S..75E}.
Our main conclusions are as follows.
\begin{enumerate}
  \item In the short shock duration model, filaments formed behind the shock start to expand/evaporate after the duration timescale of the shock, whereas the long duration model leads to star formation by forming massive filaments.
  \item The number of OB stars expected to be formed in the shock-compressed layer reaches the order of ten (i.e., massive cluster formation) when the observed peak column density exceeds 10$^{23}$ cm$^{-2}$, which is consistent with that of \citet[][]{enokiya2021PASJ...73S..75E} {(see the region with $N_{\mathrm{peak}} \sim 10^{23} \mathrm{cm^{-2}}$ in panel (b) of Figure \ref{fig:enkyDiagram_repr_abe})}. According to a simple theoretical model, such a high peak column density can be achieved only when the shock-compressed layer undergoes gravitational collapse.
  \item The massive star formation can be activated if shock compression is maintained for more than two free-fall times in the compressed layer ($t_{\mathrm{dur}}/t_{\mathrm{ff}} \gtrsim 2$). This conclusion is not significantly affected by the spatial resolution.
\end{enumerate}

\begin{acknowledgments}
We thank S. Inutsuka, K. Tokuda, and K. E. I. Tanaka for fruitful discussions. The numerical computations were carried out on XC50 system at the Center for Computational Astrophysics of National Astronomical Observatory of Japan. This work is supported by Grant-in-aids from the Ministry of Education, Culture, Sports, Science, and Technology (MEXT) of Japan (JP22J15861, 18H05436).
\end{acknowledgments}

\vspace{5mm}






\bibliography{ms}{}
\bibliographystyle{aasjournal}



\end{document}